\documentclass[aps,prl,twocolumn,floatfix]{revtex4-2}

\usepackage{amssymb}
\usepackage{natbib}
\usepackage{graphicx}
\usepackage{amsmath}
\usepackage[bookmarks = false]{hyperref}
\usepackage{bm}
\usepackage[all]{hypcap}
\usepackage{graphicx}
\usepackage{colortbl}
\usepackage{booktabs}
\usepackage{enumerate}
\usepackage{enumitem}
\usepackage{braket}
\usepackage{mathrsfs}
\usepackage{multirow}
\usepackage{upgreek}
\usepackage{textcomp}

\usepackage{soul}
\usepackage{dsfont}
\usepackage{babel}

\begin{document}

\title{Entangling two Rydberg Superatoms via Heralded Storage}

\author{Zi-Ye An$^{1,\,2,\,3\,*}$}
\author{Bo-Wei Lu$^{1,\,2,\,3\,*}$}
\author{Jun Li$^{1,\,2,\,3\,*}$}
\author{Chao-Wei Yang$^{1,\,2,\,3}$}
\author{Li Li$^{1,\,2,\,3}$}
\author{Xiao-Hui Bao$^{1,\,2,\,3}$}
\author{Jian-Wei Pan$^{1,\,2,\,3}$}

\affiliation{$^1$Hefei National Research Center for Physical Sciences at the Microscale and School of Physical Sciences, University of Science and Technology of China, Hefei 230026, China}
\affiliation{$^2$CAS Center for Excellence in Quantum Information and Quantum Physics, University of Science and Technology of China, Hefei, 230026, China}
\affiliation{$^3$Hefei National Laboratory, University of Science and Technology of China, Hefei 230088, China}
\affiliation{$^*$These authors contributed equally to this work.}

\begin{abstract}
Heralded storage of photons is crucial for advancing quantum networks. Previous realizations have primarily relied on single atoms strongly coupled to optical cavities. In this work, we present the experimental realization of heralded storage using a Rydberg superatom, a mesoscopic atomic ensemble operating in the strong blockade regime. In our approach, an input photon is initially stored in the superatom via electromagnetically induced transparency. Subsequently, a second photon is emitted conditioned on the success of the first photon's storage. Due to the collectively enhanced interaction, both the storage and the emission of the herald photon can be rather efficient in principle. As a demonstration of this technique, we use it to entangle two remote Rydberg superatoms. This method obviates the need for an intermediate node, which is commonly employed in traditional interference-based remote entanglement schemes. Our results showcase the potential of performing cavity-QED-like experiments with Rydberg superatoms. This work opens pathways for numerous applications in quantum networks and linear optical quantum computing. 
\end{abstract}

\maketitle

Quantum network~\cite{kimble_quantum_2008,wehner_quantum_2018} gives promise for a number of groundbreaking applications such as distributed quantum computing and long-distance quantum communication via quantum repeaters. A fundamental building block of a quantum network is the establishment of heralded remote entanglement between distant nodes, each equipped with matter qubits capable of storing quantum states for extended durations~\cite{azuma_quantum_2023}. To entangle remote matter qubits over long distances, an efficient matter-photon interface at each node is essential, using photons to mediate interactions between nodes. The most widely adopted approach involves creating matter-photon entanglement at each node and transmitting photons from neighboring nodes to an intermediate station, where interference and detection enable heralded entanglement. This scheme has been successfully implemented in achieving memory-based entanglement over fiber links spanning tens of kilometers~\cite{yu_entanglement_2020,van_leent_entangling_2022}, even in field-deployed scenarios~\cite{liu_creation_2024,stolk_metropolitan-scale_2024}.

For practical deployment, however, it is desirable to reduce resource overhead, particularly the number of physically separated nodes. If heralded storage of photons is available, one can eliminate the need for an intermediate interference. In this way, matter-photon entanglement is firstly created at one node, and the photon is transmitted to a second node, where its quantum state is transferred to a matter qubit in a heralded manner. A straightforward way to achieve this is through heralded absorption of a single photon by a single atom. However, this approach is rather inefficient in free space~\cite{kurz_experimental_2014}, even with high-numerical-aperture lenses. Applying high-finesse optical resonators helps to facilitate the input photon's storage and the herald photon's emission \cite{brekenfeld2020quantum}. By placing the single atom in a cavity and operating in the strong coupling regime, heralded matter-photon gates can also be realized in an efficient way~\cite{kalb_heralded_2015,bhaskar_experimental_2020}. This technique has been successfully demonstrated for entangling remote single neutral atoms~\cite{daiss_quantum-logic_2021} and silicon-vacancy centers in diamonds~\cite{knaut_entanglement_2024}.

Compared to single atoms, atomic ensembles exhibit collective atom-photon interactions, making them ideal candidates for quantum memories. High storage efficiencies have been demonstrated using schemes such as electromagnetically induced transparency (EIT)~\cite{fleischhauer_electromagnetically_2005} and Raman protocols~\cite{nunn_mapping_2007}. However, these schemes lack the ability to herald successful storage events. A pioneering experiment using atomic ensembles detected a Raman-scattered single photon as a heralding signal~\cite{tanji_heralded_2009}, but this approach proved to be highly inefficient. 

\begin{figure*}[htp]
	\centering
	\includegraphics[width=0.9\textwidth]{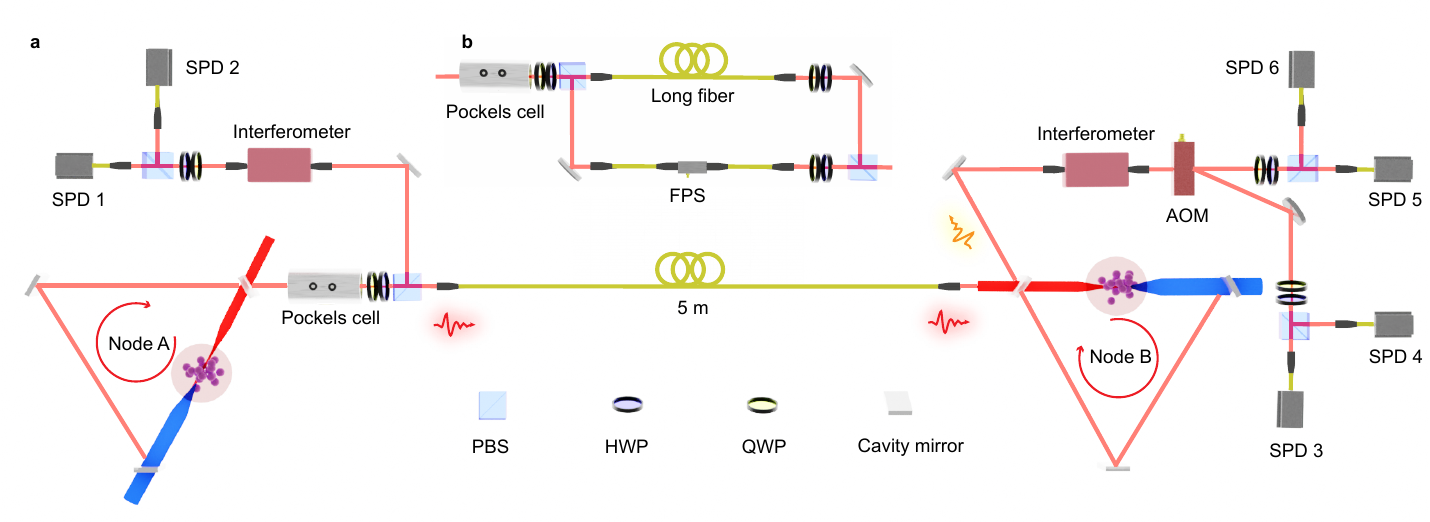}
	 \caption{The experimental setup. (a) Layout of our experimental setup. Node A and Node B are both laser-cooled $^{87}\mathrm{Rb}$ ensembles with a low-finesse cavity, where Node A serves either as a single-photon source or an atom-photon entanglement source, and Node B serves as a heralded quantum memory. The single photon generated from Node A is either sent to Node B or measured locally, which is controlled by the Pockels cell that switches the polarization of the photon. In node B, the photon to be stored is incident through the cavity mirror. The herald photon and the photon retrieved are directed to SPD3/SPD4 and SPD5/SPD6, respectively, by an acousto-optic modulator (AOM). The cavities are used to enhance the retrieval efficiency of the photons. (b) The Mach-Zehnder interferometer used for converting the time-bin-encoded photons to polarization-encoded photons. The Pockels cell directs the early mode and the late mode to the long fibre and the short fibre, respectively. The fiber phase shifter (FPS) is mounted on the short fiber for phase stabilization.}
	\label{fig:Fig1}
\end{figure*}

In this paper, we report an experiment leveraging Rydberg interactions~\cite{saffman_quantum_2010} in a mesoscopic atomic ensemble—referred to as a Rydberg superatom \cite{Kumlin_2023}—to achieve heralded photon storage. A single photon encoded in the time-bin basis is first stored in the Rydberg superatom using the EIT mechanism\cite{fleischhauer_electromagnetically_2005,PhysRevLett.105.193603,PhysRevLett.107.213601,peyronel2012quantum,PhysRevLett.112.073901,PhysRevLett.113.053601,PhysRevLett.113.053602,tiarks2019photon,vaneecloo2022intracavity,stolz2022quantum}. Subsequently, by applying a tailored `read-and-patch' pulse sequence developed from our previous works~\cite{sun_deterministic_2022,yang_sequential_2022}, a second photon is emitted, conditioned on the successful storage of the first photon in the superatom. We measure the storage fidelities thoroughly. We further demonstrate the utility of this heralded storage mechanism by entangling two remote Rydberg superatoms. The entanglement is verified, showcasing the feasibility of this approach for simplifying quantum network architectures. Beyond quantum networks, the demonstrated heralded storage could find applications in linear optical quantum computing~\cite{obrien_optical_2007} and security testing for quantum key distribution systems~\cite{xu_secure_2020}, etc. 

Our experimental setup is depicted in Fig.~\ref{fig:Fig1}a. Node A and Node B are two $^{87}\mathrm{Rb}$  atomic ensembles with similar experimental conditions. Atoms in both ensembles are laser-cooled by magneto-optical trap (MOT) and optical molasses, which are then trapped in far red-detuned optical trap with temperature of 6~$\mathrm{\upmu K}$ and optical depth of 2. The atoms are prepared in $|g \rangle = | 5 S_{1/2}, F=2, m_{F} = +2 \rangle$ by optical pumping, and a moderate biased magnetic field is applied to lift the degeneracy of the Zeeman sublevels. The 795 nm lasers (red cylinder in Fig.~\ref{fig:Fig1}a) at both Node A and Node B couple the ground state $|g \rangle$ and the intermediate state $|e_{1}\rangle  = | 5 P_{1/2}, F=2, m_{F}=+1 \rangle $ with a detuning of about 60 MHz. The 475 nm laser (blue cylinder in Fig.~\ref{fig:Fig1}a) couples the $|e_{1} \rangle$ and the Rydberg state $|r \rangle = | 91 S_{1/2} \rangle$ to address the two-photon excitation from the ground state to the Rydberg state together with the 795 nm beam. The 475 nm laser beam at Node B which is on resonance with the state $|e_{2}\rangle  = | 5 P_{1/2}, F=1, m_{F} = +1 \rangle $ also serves as the control beam of the EIT storage process. In addition, the low finesse cavities are applied in both nodes to enhance the readout process\cite{yang_sequential_2022}. The cavities have a finesse of $\simeq19$ and resonate with both the 795 nm excitation beams and the output single photons.

\begin{figure*}[htp]
	\centering
	\includegraphics[width=0.9\textwidth]{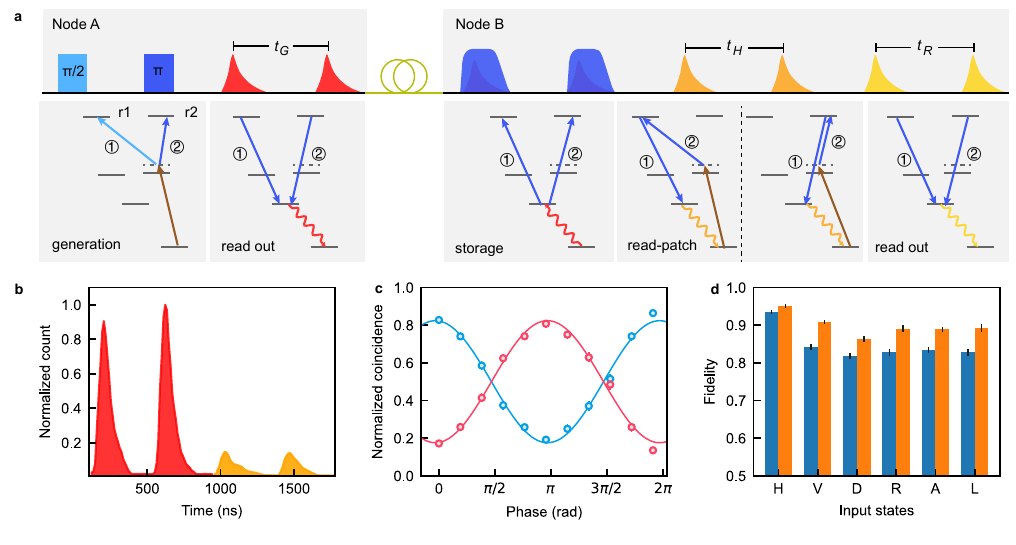}
	 \caption{The EIT storage and state tomography. (a) Time sequence of EIT storage of a single qubit and the corresponding energy levels. The blue and red arrows in the energy diagrams refer to the 475 nm pulses and 795 nm pulses, respectively. The  wavy arrows in the diagram refer to the single photon stored into or emitted from the ensemble.  Time intervals between time bins of the photon generated at Node A ($t_{G}$),  of the heralding qubit ($t_{H}$) and of the retrieval qubit ($t_{R}$) are all 425 ns. (b) The profile of input time-bin photon (red) and the profile of the photon after the storing and retrieving (orange). (c) The verification of entanglement between the herald photon and the retrieved photon at Node B.  The light blue circles are the sum of coincidences of $|D\rangle|D^\prime\rangle$ and $|A\rangle|A^\prime\rangle$, while the light red circles are the sum of coincidences of $|D\rangle|A^\prime\rangle$ and $|A\rangle|D^\prime\rangle$. The coincidences are normalized by the total coincidences of each phase point. The solid lines are the fitting curves corresponding to the normalized coincidences. (d) The state fidelity of six input states. The blue bars are the results derived from raw data and the orange bars are the results deducting dark counts.}
	\label{fig:Fig2}
\end{figure*}

The single photon is generated by Node A via two-photon Rydberg excitation. A $\pi/2$ pulse coupling $|g\rangle$ to $|r_{1}\rangle$ followed by a $\pi$ pulse coupling $|g\rangle$ to $|r_{2}\rangle$ is applied (see Fig.~\ref{fig:Fig2}a),where $|r_{1}\rangle$ and  $|r_{2}\rangle$ are the Zeeman sublevels of $|r\rangle$. Due to the Rydberg blockade effect, there is only one atom excited to the Rydberg state inside the blockade radius\cite{2012Strongly,PhysRevLett.112.073901,sun_deterministic_2022,yang_sequential_2022}, leading the atomic ensemble to be the superposition between $|R_{1}\rangle$ and $|R_{2}\rangle$, where $|R_{1}\rangle$ and $|R_{2}\rangle$ refer to the the collective excitation on the Zeeman sublevel of $|r_{1}\rangle$ and $|r_{2}\rangle$, respectively. We then apply two consecutive read-out pulses coupling $|r_{1}\rangle$ and $|r_{2}\rangle$ with the intermediate state, which generates the early mode $|t_{E}\rangle$ and the late mode $|t_{L}\rangle$ of the time-bin photonic qubit, respectively. The time-bin photonic state, whose time interval between early mode and late mode is 425 ns, is then transmitted to Node B via a fiber and stored via Rydberg EIT process with a 475 nm control beam covering the incoming photons. The profile of the input photonic qubit and the output qubit is depicted in Fig.~\ref{fig:Fig2}b and we reach an overall storage and read-out efficiency of $\eta_{\mathrm{sr}} = 16.4\%$.  After storing for about 670 ns, we perform two consecutive `read-and-patch' processes at Node B, which generates the time-bin qubit for heralding, and the detection of the herald qubit signals the successful storage of the input photon.

The `read-and-patch' process after storing creates the atom-photon entanglement
\begin{equation}
|\psi_{\mathrm{ap}}\rangle = \frac{1}{\sqrt{2}} (|R_{1}\rangle |t_{E}\rangle+ e^{\mathrm{i} \theta_{0}}|R_{2}\rangle |t_{L}\rangle).
\end{equation}
To verify the entanglement, we retrieve the atomic qubit and convert the time-bin qubit to polarization-encoded one for further measurement. Thus we apply two read pulses again and convert $|R_{1}\rangle$, $|R_{2}\rangle$ to $|t^\prime_{E}\rangle$ and $|t^\prime_{L}\rangle$, respectively, establishing a photon-photon entanglement between the herald photon and the retrieved photon. \newline

The conversion from time-bin qubit to polarization qubit is realized by an interferometer with a long arm of 90 m and a short arm of 5 m, which matches the time interval of the time-bin qubit (see Fig.~\ref{fig:Fig1}b). The interferometer maps early mode $|t_{E}\rangle$ to $V$ polarization and late mode $|t_{L}\rangle$ to $H$ polarization, and is locked by a locking beam injected into the other port of the PBS with the feedback signal imposed on the fiber phase shifter (FPS) to keep the relative phase $\theta_{0}$ stable. The paths of herald qubit and retrieval qubit after emitting from the cavity are toggled by an AOM so that we are able to perform arbitrary rotational operations to the retrieved photons independently. We verify the entanglement by setting waveplates in both heralding path and retrieved path to $|+\rangle/|-\rangle$ basis and performing the correlation measurement between click events of SPD3/SPD4 and SPD5/SPD6. By sweeping the phase of read pulse, we expect sinusoidal oscillations of the coincidence counts between the herald photons and the retrieved photons, which is shown in Fig.~\ref{fig:Fig2}c. The fitted curve yields the visibility of $V_{1} = 64.7\% \pm 1.7\%$. This result, together with the visibility in $|H\rangle/|V\rangle$ basis at a fixed phase of $V_{0} = 79.9\% \pm 1.4\%$, give rise to the entanglement fidelity of $F_{e} = (1 + V_{0} + 2V_{1})/4 = 77.3\% \pm 0.9\%$.

\begin{figure}[htp]
	\centering
	\includegraphics[width=0.45\textwidth]{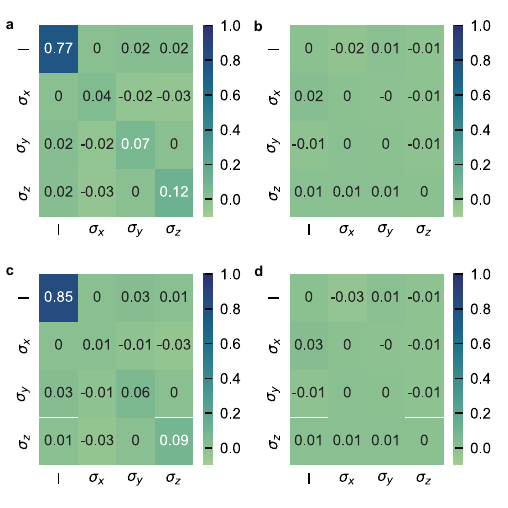}
	 \caption{Process tomography of the memory. (a) and (b) The real part and the imaginary part of elements of the process matrix $\chi$ of the raw data, respectively. (c) and (d) The real part and the imaginary part of elements of process matrix which deducts the dark counts, respectively.
	 }
	\label{fig:Fig3}
\end{figure}

To quantify the performance of our memory,  we prepare six input states $|t_{E}\rangle / |t_{L}\rangle$, $|D \rangle / |A \rangle$ ($1/\sqrt{2}(|t_{E}\rangle \pm |t_{L}\rangle)$) and $|R \rangle / |L \rangle$ ($1/\sqrt{2}(|t_{E}\rangle \pm \mathrm{i} |t_{L}\rangle)$) at Node A and perform state tomography for all corresponding retrieved photon states after heralded storage. We keep the heralding path on $|+\rangle/|-\rangle$ basis and perform three rotational operations $I$, $R_{y}(-\pi/2)$ and $R_{x}(\pi/2)$ to retrieved photons to reconstruct the density matrix of the output states.  The results are shown in Fig 2d. The fidelity is defined as $F_{s} = \mathrm{Tr}(\rho_{\mathrm{out}} |\psi_{\mathrm{in}} \rangle \langle \psi_{\mathrm{in}} |)$, where $| \psi_{\mathrm{in}} \rangle$ denotes the input states and $\rho_{\mathrm{out}}$ denotes the reconstructed density matrix of retrieved photon states. We get the raw average state fidelity of $F_{\mathrm{sr}} = 84.8\% \pm 0.3\%$ and deducted average state fidelity of $F_{\mathrm{sd}} = 89.9\% \pm 0.3\% $ which eliminates the effect of accidental coincidences between the dark counts of SPDs and the retrieved photons. 

The storage of a qubit in quantum memory can be regarded as a quantum process, which is usually characterized by the process matrix $\chi$. The process matrix of an ideal quantum memory storing a single qubit $\chi_{0}$ is a matrix with $\chi_{0}(1, 1)=1$ and all other elements to be zero. To reconstruct the process $\chi$ and quantify the fidelity of the storage process, we perform process tomography by choosing the coincidence counts acquired in state tomography measurement of $|t_{E}\rangle$, $|t_{L}\rangle$, $|D\rangle$ and $|R\rangle$ and computing the quantum process matrix using maximum-likelihood method~\cite{greenbaum2015introduction,beterov2016simulated,brekenfeld2020quantum}. The process tomography fidelity is defined as $F_{p} = 1 - \frac{1}{2}\mathrm{Tr}\sqrt{(\chi_{0} - \chi)^\dagger(\chi_{0} - \chi)}$. We get fidelity of $F_{\mathrm{pr}} = 76.4\% \pm 0.9\%$ for raw data and $F_{\mathrm{pd}} = 83.4\% \pm 0.8\% $ for data deducting dark counts of SPDs, both of which are beyond the classical limit of $69\%$, indicating the quantum nature of our storage process. 

Furthermore, we define the herald efficiency as $\eta_{h} = {\eta}'_{\mathrm{sr}} \eta_{t} \eta_{d}$, where ${\eta}'_{\mathrm{sr}} = 6.8\%$ is the storing-and-retrieving efficiency of a photon throughout the `read-and patch' process, $\eta_{t} = 48.1\%$ is the transmission efficiency, and $\eta_{d} = 65.6\%$ is the SPD efficiency. These results lead to a herald efficiency of $\eta_{h} = 2.1\%$. The decay of ${\eta}'_{\mathrm{sr}}$ compared to ${\eta}_{\mathrm{sr}}$ is mainly due to motion-induced dephasing during storage \cite{yang_sequential_2022}. At present, the lifetime of the Rydberg spin-wave is around 1.4 $\mu s$. In the future, feasible improvements including implementing optical lattice and transferring the Rydberg atoms to ground states to increase the storage lifetime will be realistic.

\begin{figure}[htp]
	\centering
	\includegraphics[width=0.45\textwidth]{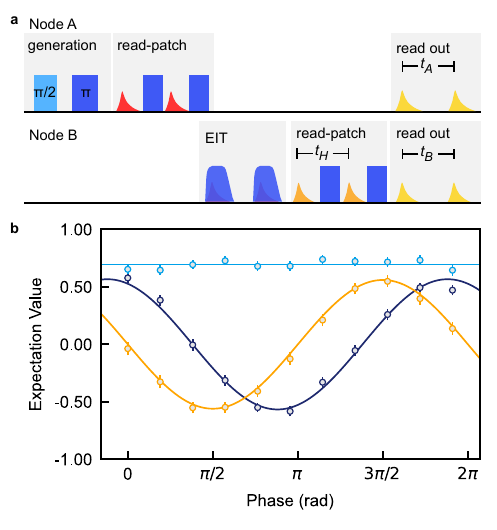}
	 \caption{Entanglement between two nodes. (a) Time sequence of two node entanglement. Compared to the time sequence of heralded storage, a `read-and-patch' process followed by the single photon generation and the read-out process are performed at Node A. (b) The results of entanglement verification. The purple dots, orange dots and blue dots refer to the measured results of  $\langle XX\rangle$,  $\langle YY\rangle$ and  $\langle ZZ\rangle$, respectively, via changing the phase of Node B. The solid lines are the fitting curves of the expectation values. }
	\label{fig:Fig4}
\end{figure}

An important application of the heralded quantum memory is to generate entanglement between two quantum memories without an intermediate node. In our setup, this entanglement can be generated and verified by slightly modifying the time sequence at Node A. As is shown in Fig.~\ref{fig:Fig4}a, after the superposition state of Rydberg collective excitations is established, we also apply the `read-and-patch' procedure to both $|r_{1}\rangle$ and $|r_{2}\rangle$ at Node A instead of direct reading out, which emits the flying qubit and preserve the superposition of the collective Rydberg excitation. Thus the atom-photon entanglement is generated initially at Node A. The entanglement between two remote nodes is then established after the storing process together with the two consecutive `read-and-patch' processes at Node B, conditioned on the successful detection of the herald photon on a superposition basis, which is
\begin{equation}
|\psi^{\mathrm{AB}}\rangle = \frac{1}{\sqrt{2}} (|R^A_{\mathrm{1}} \rangle |R^B_{\mathrm{1}}\rangle + e^{\mathrm{i}\theta_1}|R^A_{\mathrm{2}}\rangle |R^B_{\mathrm{2}}\rangle).
\end{equation}

The entanglement is verified by applying two read pulses at both Node A and Node B to retrieve the atomic states, leading to the photon-photon entanglement $|\psi_{\mathrm{pp}}\rangle = 1/\sqrt{2}(|t^A_E\rangle |t^B_E\rangle + e^{\mathrm{i} \theta_2}|t^A_L\rangle |t^B_L\rangle)$, which are then converted to polarization qubits via interferometers located on both nodes.

Fig.~\ref{fig:Fig4}b shows the verification of the entanglement between two nodes. Like entanglement measurement in a single node, we sweep the phase of the reading pulses at Node B and collect the coincidence events between clicks of SPD1/SPD2 at Node A and SPD5/SPD6 at Node B, conditioned on those of SPD3/SPD4 as heralding signals. The entanglement fidelity is defined as $F_e=(1+\langle XX\rangle-\langle YY\rangle+\langle ZZ\rangle)/4$, where X, Y and Z denote the standard Pauli matrix of $\sigma_X$, $\sigma_Y$ and $\sigma_Z$, respectively. We fit the curves and get the expectation values of $\langle XX\rangle=56.7\% \pm 2.0\% $, $\langle YY\rangle= -56.0\% \pm 1.3\%$ and $\langle ZZ\rangle=69.4\% \pm 1.1\%$, which yields the entanglement fidelity of $F_e=70.5\%\pm 0.6\%$. This result clearly exceeds the classical limit bound of $F_{e} > 50\%$. 

The herald efficiency and measurement fidelity of our quantum memory is currently limited by the relatively low efficiency of photon storing and retrieving. This limitation is not difficult to relieve by moderately improving the interaction strength of the photon-atom interface with better cavities and larger optical depth of atoms, as well as applying high-performance superconducting nanowire single-photon detectors (SNSPD). On the other hand, finely optimizing the profile of the control beams also helps to enhance the storage efficiency, while the fidelity requires an optimal Rydberg $\pi$ pulse.

In summary, we realize the heralded storage of a single qubit in a Rydberg atomic ensemble and demonstrate the generation of entanglement between two nodes. By utilizing the Rydberg blockade effect, we generate herald photons via a read-and-patch process. After the heralding signal is detected, the entanglement between two superatoms is established immediately, and is then verified via coincidence measurements of retrieved photons. We show that with the heralded quantum memory in our work, entanglement between two nodes can be generated without an intermediate node. With further developments, the demonstrated heralded storage and two-node entanglement may become essential elements for ensemble-based quantum repeater and quantum networks.

\section{Acknowledgment}
This research was supported by the Innovation Program for Quantum Science and Technology (No. 2021ZD0301101), National Natural Science Foundation of China (No.~12274394, No.~12274399), the Chinese Academy of Sciences, USTC Research Funds of the Double First-Class Initiative (YD9990002012), and China Postdoctoral Science Foundation (BX20230105, 2023M730901).

\bibliography{myref}

\begin{thebibliography}{34}%
\makeatletter
\providecommand \@ifxundefined [1]{%
 \@ifx{#1\undefined}
}%
\providecommand \@ifnum [1]{%
 \ifnum #1\expandafter \@firstoftwo
 \else \expandafter \@secondoftwo
 \fi
}%
\providecommand \@ifx [1]{%
 \ifx #1\expandafter \@firstoftwo
 \else \expandafter \@secondoftwo
 \fi
}%
\providecommand \natexlab [1]{#1}%
\providecommand \enquote  [1]{``#1''}%
\providecommand \bibnamefont  [1]{#1}%
\providecommand \bibfnamefont [1]{#1}%
\providecommand \citenamefont [1]{#1}%
\providecommand \href@noop [0]{\@secondoftwo}%
\providecommand \href [0]{\begingroup \@sanitize@url \@href}%
\providecommand \@href[1]{\@@startlink{#1}\@@href}%
\providecommand \@@href[1]{\endgroup#1\@@endlink}%
\providecommand \@sanitize@url [0]{\catcode `\\12\catcode `\$12\catcode
  `\&12\catcode `\#12\catcode `\^12\catcode `\_12\catcode `\%12\relax}%
\providecommand \@@startlink[1]{}%
\providecommand \@@endlink[0]{}%
\providecommand \url  [0]{\begingroup\@sanitize@url \@url }%
\providecommand \@url [1]{\endgroup\@href {#1}{\urlprefix }}%
\providecommand \urlprefix  [0]{URL }%
\providecommand \Eprint [0]{\href }%
\providecommand \doibase [0]{https://doi.org/}%
\providecommand \selectlanguage [0]{\@gobble}%
\providecommand \bibinfo  [0]{\@secondoftwo}%
\providecommand \bibfield  [0]{\@secondoftwo}%
\providecommand \translation [1]{[#1]}%
\providecommand \BibitemOpen [0]{}%
\providecommand \bibitemStop [0]{}%
\providecommand \bibitemNoStop [0]{.\EOS\space}%
\providecommand \EOS [0]{\spacefactor3000\relax}%
\providecommand \BibitemShut  [1]{\csname bibitem#1\endcsname}%
\let\auto@bib@innerbib\@empty
\bibitem [{\citenamefont {Kimble}(2008)}]{kimble_quantum_2008}%
  \BibitemOpen
  \bibfield  {author} {\bibinfo {author} {\bibfnamefont {H.~J.}\ \bibnamefont
  {Kimble}},\ }\bibfield  {title} {\bibinfo {title} {The quantum internet},\
  }\href@noop {} {\bibfield  {journal} {\bibinfo  {journal} {Nature}\ }\textbf
  {\bibinfo {volume} {453}},\ \bibinfo {pages} {1023} (\bibinfo {year}
  {2008})}\BibitemShut {NoStop}%
\bibitem [{\citenamefont {Wehner}\ \emph {et~al.}(2018)\citenamefont {Wehner},
  \citenamefont {Elkouss},\ and\ \citenamefont {Hanson}}]{wehner_quantum_2018}%
  \BibitemOpen
  \bibfield  {author} {\bibinfo {author} {\bibfnamefont {S.}~\bibnamefont
  {Wehner}}, \bibinfo {author} {\bibfnamefont {D.}~\bibnamefont {Elkouss}},\
  and\ \bibinfo {author} {\bibfnamefont {R.}~\bibnamefont {Hanson}},\
  }\bibfield  {title} {\bibinfo {title} {Quantum internet: {A} vision for the
  road ahead},\ }\href {https://doi.org/10.1126/science.aam9288} {\bibfield
  {journal} {\bibinfo  {journal} {Science}\ }\textbf {\bibinfo {volume}
  {362}},\ \bibinfo {pages} {eaam9288} (\bibinfo {year} {2018})}\BibitemShut
  {NoStop}%
\bibitem [{\citenamefont {Azuma}\ \emph {et~al.}(2023)\citenamefont {Azuma},
  \citenamefont {Economou}, \citenamefont {Elkouss}, \citenamefont {Hilaire},
  \citenamefont {Jiang}, \citenamefont {Lo},\ and\ \citenamefont
  {Tzitrin}}]{azuma_quantum_2023}%
  \BibitemOpen
  \bibfield  {author} {\bibinfo {author} {\bibfnamefont {K.}~\bibnamefont
  {Azuma}}, \bibinfo {author} {\bibfnamefont {S.~E.}\ \bibnamefont {Economou}},
  \bibinfo {author} {\bibfnamefont {D.}~\bibnamefont {Elkouss}}, \bibinfo
  {author} {\bibfnamefont {P.}~\bibnamefont {Hilaire}}, \bibinfo {author}
  {\bibfnamefont {L.}~\bibnamefont {Jiang}}, \bibinfo {author} {\bibfnamefont
  {H.-K.}\ \bibnamefont {Lo}},\ and\ \bibinfo {author} {\bibfnamefont
  {I.}~\bibnamefont {Tzitrin}},\ }\bibfield  {title} {\bibinfo {title} {Quantum
  repeaters: {From} quantum networks to the quantum internet},\ }\href
  {https://doi.org/10.1103/RevModPhys.95.045006} {\bibfield  {journal}
  {\bibinfo  {journal} {Reviews of Modern Physics}\ }\textbf {\bibinfo {volume}
  {95}},\ \bibinfo {pages} {045006} (\bibinfo {year} {2023})}\BibitemShut
  {NoStop}%
\bibitem [{\citenamefont {Yu}\ \emph {et~al.}(2020)\citenamefont {Yu},
  \citenamefont {Ma}, \citenamefont {Luo}, \citenamefont {Jing}, \citenamefont
  {Sun}, \citenamefont {Fang}, \citenamefont {Yang}, \citenamefont {Liu},
  \citenamefont {Zheng}, \citenamefont {Xie}, \citenamefont {Zhang},
  \citenamefont {You}, \citenamefont {Wang}, \citenamefont {Chen},
  \citenamefont {Zhang}, \citenamefont {Bao},\ and\ \citenamefont
  {Pan}}]{yu_entanglement_2020}%
  \BibitemOpen
  \bibfield  {author} {\bibinfo {author} {\bibfnamefont {Y.}~\bibnamefont
  {Yu}}, \bibinfo {author} {\bibfnamefont {F.}~\bibnamefont {Ma}}, \bibinfo
  {author} {\bibfnamefont {X.-Y.}\ \bibnamefont {Luo}}, \bibinfo {author}
  {\bibfnamefont {B.}~\bibnamefont {Jing}}, \bibinfo {author} {\bibfnamefont
  {P.-F.}\ \bibnamefont {Sun}}, \bibinfo {author} {\bibfnamefont {R.-Z.}\
  \bibnamefont {Fang}}, \bibinfo {author} {\bibfnamefont {C.-W.}\ \bibnamefont
  {Yang}}, \bibinfo {author} {\bibfnamefont {H.}~\bibnamefont {Liu}}, \bibinfo
  {author} {\bibfnamefont {M.-Y.}\ \bibnamefont {Zheng}}, \bibinfo {author}
  {\bibfnamefont {X.-P.}\ \bibnamefont {Xie}}, \bibinfo {author} {\bibfnamefont
  {W.-J.}\ \bibnamefont {Zhang}}, \bibinfo {author} {\bibfnamefont {L.-X.}\
  \bibnamefont {You}}, \bibinfo {author} {\bibfnamefont {Z.}~\bibnamefont
  {Wang}}, \bibinfo {author} {\bibfnamefont {T.-Y.}\ \bibnamefont {Chen}},
  \bibinfo {author} {\bibfnamefont {Q.}~\bibnamefont {Zhang}}, \bibinfo
  {author} {\bibfnamefont {X.-H.}\ \bibnamefont {Bao}},\ and\ \bibinfo {author}
  {\bibfnamefont {J.-W.}\ \bibnamefont {Pan}},\ }\bibfield  {title} {\bibinfo
  {title} {Entanglement of two quantum memories via fibres over dozens of
  kilometres},\ }\href {https://doi.org/10.1038/s41586-020-1976-7} {\bibfield
  {journal} {\bibinfo  {journal} {Nature}\ }\textbf {\bibinfo {volume} {578}},\
  \bibinfo {pages} {240} (\bibinfo {year} {2020})}\BibitemShut {NoStop}%
\bibitem [{\citenamefont {van Leent}\ \emph {et~al.}(2022)\citenamefont {van
  Leent}, \citenamefont {Bock}, \citenamefont {Fertig}, \citenamefont
  {Garthoff}, \citenamefont {Eppelt}, \citenamefont {Zhou}, \citenamefont
  {Malik}, \citenamefont {Seubert}, \citenamefont {Bauer}, \citenamefont
  {Rosenfeld}, \citenamefont {Zhang}, \citenamefont {Becher},\ and\
  \citenamefont {Weinfurter}}]{van_leent_entangling_2022}%
  \BibitemOpen
  \bibfield  {author} {\bibinfo {author} {\bibfnamefont {T.}~\bibnamefont {van
  Leent}}, \bibinfo {author} {\bibfnamefont {M.}~\bibnamefont {Bock}}, \bibinfo
  {author} {\bibfnamefont {F.}~\bibnamefont {Fertig}}, \bibinfo {author}
  {\bibfnamefont {R.}~\bibnamefont {Garthoff}}, \bibinfo {author}
  {\bibfnamefont {S.}~\bibnamefont {Eppelt}}, \bibinfo {author} {\bibfnamefont
  {Y.}~\bibnamefont {Zhou}}, \bibinfo {author} {\bibfnamefont {P.}~\bibnamefont
  {Malik}}, \bibinfo {author} {\bibfnamefont {M.}~\bibnamefont {Seubert}},
  \bibinfo {author} {\bibfnamefont {T.}~\bibnamefont {Bauer}}, \bibinfo
  {author} {\bibfnamefont {W.}~\bibnamefont {Rosenfeld}}, \bibinfo {author}
  {\bibfnamefont {W.}~\bibnamefont {Zhang}}, \bibinfo {author} {\bibfnamefont
  {C.}~\bibnamefont {Becher}},\ and\ \bibinfo {author} {\bibfnamefont
  {H.}~\bibnamefont {Weinfurter}},\ }\bibfield  {title} {\bibinfo {title}
  {Entangling single atoms over 33 km telecom fibre},\ }\href
  {https://doi.org/10.1038/s41586-022-04764-4} {\bibfield  {journal} {\bibinfo
  {journal} {Nature}\ }\textbf {\bibinfo {volume} {607}},\ \bibinfo {pages}
  {69} (\bibinfo {year} {2022})}\BibitemShut {NoStop}%
\bibitem [{\citenamefont {Liu}\ \emph {et~al.}(2024)\citenamefont {Liu},
  \citenamefont {Luo}, \citenamefont {Yu}, \citenamefont {Wang}, \citenamefont
  {Wang}, \citenamefont {Hu}, \citenamefont {Li}, \citenamefont {Zheng},
  \citenamefont {Yao}, \citenamefont {Yan}, \citenamefont {Teng}, \citenamefont
  {Jiang}, \citenamefont {Liu}, \citenamefont {Xie}, \citenamefont {Zhang},
  \citenamefont {Mao}, \citenamefont {Jiang}, \citenamefont {Zhang},
  \citenamefont {Bao},\ and\ \citenamefont {Pan}}]{liu_creation_2024}%
  \BibitemOpen
  \bibfield  {author} {\bibinfo {author} {\bibfnamefont {J.-L.}\ \bibnamefont
  {Liu}}, \bibinfo {author} {\bibfnamefont {X.-Y.}\ \bibnamefont {Luo}},
  \bibinfo {author} {\bibfnamefont {Y.}~\bibnamefont {Yu}}, \bibinfo {author}
  {\bibfnamefont {C.-Y.}\ \bibnamefont {Wang}}, \bibinfo {author}
  {\bibfnamefont {B.}~\bibnamefont {Wang}}, \bibinfo {author} {\bibfnamefont
  {Y.}~\bibnamefont {Hu}}, \bibinfo {author} {\bibfnamefont {J.}~\bibnamefont
  {Li}}, \bibinfo {author} {\bibfnamefont {M.-Y.}\ \bibnamefont {Zheng}},
  \bibinfo {author} {\bibfnamefont {B.}~\bibnamefont {Yao}}, \bibinfo {author}
  {\bibfnamefont {Z.}~\bibnamefont {Yan}}, \bibinfo {author} {\bibfnamefont
  {D.}~\bibnamefont {Teng}}, \bibinfo {author} {\bibfnamefont {J.-W.}\
  \bibnamefont {Jiang}}, \bibinfo {author} {\bibfnamefont {X.-B.}\ \bibnamefont
  {Liu}}, \bibinfo {author} {\bibfnamefont {X.-P.}\ \bibnamefont {Xie}},
  \bibinfo {author} {\bibfnamefont {J.}~\bibnamefont {Zhang}}, \bibinfo
  {author} {\bibfnamefont {Q.-H.}\ \bibnamefont {Mao}}, \bibinfo {author}
  {\bibfnamefont {X.}~\bibnamefont {Jiang}}, \bibinfo {author} {\bibfnamefont
  {Q.}~\bibnamefont {Zhang}}, \bibinfo {author} {\bibfnamefont {X.-H.}\
  \bibnamefont {Bao}},\ and\ \bibinfo {author} {\bibfnamefont {J.-W.}\
  \bibnamefont {Pan}},\ }\bibfield  {title} {\bibinfo {title} {Creation of
  memory memory entanglement in a metropolitan quantum network},\ }\href
  {https://doi.org/10.1038/s41586-024-07308-0} {\bibfield  {journal} {\bibinfo
  {journal} {Nature}\ }\textbf {\bibinfo {volume} {629}},\ \bibinfo {pages}
  {579} (\bibinfo {year} {2024})}\BibitemShut {NoStop}%
\bibitem [{\citenamefont {Stolk}\ \emph {et~al.}(2024)\citenamefont {Stolk},
  \citenamefont {Van Der~Enden}, \citenamefont {Slater}, \citenamefont
  {Te~Raa-Derckx}, \citenamefont {Botma}, \citenamefont {Van~Rantwijk},
  \citenamefont {Biemond}, \citenamefont {Hagen}, \citenamefont {Herfst},
  \citenamefont {Koek}, \citenamefont {Meskers}, \citenamefont {Vollmer},
  \citenamefont {Van~Zwet}, \citenamefont {Markham}, \citenamefont {Edmonds},
  \citenamefont {Geus}, \citenamefont {Elsen}, \citenamefont {Jungbluth},
  \citenamefont {Haefner}, \citenamefont {Tresp}, \citenamefont {Stuhler},
  \citenamefont {Ritter},\ and\ \citenamefont
  {Hanson}}]{stolk_metropolitan-scale_2024}%
  \BibitemOpen
  \bibfield  {author} {\bibinfo {author} {\bibfnamefont {A.~J.}\ \bibnamefont
  {Stolk}}, \bibinfo {author} {\bibfnamefont {K.~L.}\ \bibnamefont {Van
  Der~Enden}}, \bibinfo {author} {\bibfnamefont {M.-C.}\ \bibnamefont
  {Slater}}, \bibinfo {author} {\bibfnamefont {I.}~\bibnamefont
  {Te~Raa-Derckx}}, \bibinfo {author} {\bibfnamefont {P.}~\bibnamefont
  {Botma}}, \bibinfo {author} {\bibfnamefont {J.}~\bibnamefont {Van~Rantwijk}},
  \bibinfo {author} {\bibfnamefont {J.~J.~B.}\ \bibnamefont {Biemond}},
  \bibinfo {author} {\bibfnamefont {R.~A.~J.}\ \bibnamefont {Hagen}}, \bibinfo
  {author} {\bibfnamefont {R.~W.}\ \bibnamefont {Herfst}}, \bibinfo {author}
  {\bibfnamefont {W.~D.}\ \bibnamefont {Koek}}, \bibinfo {author}
  {\bibfnamefont {A.~J.~H.}\ \bibnamefont {Meskers}}, \bibinfo {author}
  {\bibfnamefont {R.}~\bibnamefont {Vollmer}}, \bibinfo {author} {\bibfnamefont
  {E.~J.}\ \bibnamefont {Van~Zwet}}, \bibinfo {author} {\bibfnamefont
  {M.}~\bibnamefont {Markham}}, \bibinfo {author} {\bibfnamefont {A.~M.}\
  \bibnamefont {Edmonds}}, \bibinfo {author} {\bibfnamefont {J.~F.}\
  \bibnamefont {Geus}}, \bibinfo {author} {\bibfnamefont {F.}~\bibnamefont
  {Elsen}}, \bibinfo {author} {\bibfnamefont {B.}~\bibnamefont {Jungbluth}},
  \bibinfo {author} {\bibfnamefont {C.}~\bibnamefont {Haefner}}, \bibinfo
  {author} {\bibfnamefont {C.}~\bibnamefont {Tresp}}, \bibinfo {author}
  {\bibfnamefont {J.}~\bibnamefont {Stuhler}}, \bibinfo {author} {\bibfnamefont
  {S.}~\bibnamefont {Ritter}},\ and\ \bibinfo {author} {\bibfnamefont
  {R.}~\bibnamefont {Hanson}},\ }\bibfield  {title} {\bibinfo {title}
  {Metropolitan-scale heralded entanglement of solid-state qubits},\ }\href
  {https://doi.org/10.1126/sciadv.adp6442} {\bibfield  {journal} {\bibinfo
  {journal} {Science Advances}\ }\textbf {\bibinfo {volume} {10}},\ \bibinfo
  {pages} {eadp6442} (\bibinfo {year} {2024})}\BibitemShut {NoStop}%
\bibitem [{\citenamefont {Kurz}\ \emph {et~al.}(2014)\citenamefont {Kurz},
  \citenamefont {Schug}, \citenamefont {Eich}, \citenamefont {Huwer},
  \citenamefont {Müller},\ and\ \citenamefont
  {Eschner}}]{kurz_experimental_2014}%
  \BibitemOpen
  \bibfield  {author} {\bibinfo {author} {\bibfnamefont {C.}~\bibnamefont
  {Kurz}}, \bibinfo {author} {\bibfnamefont {M.}~\bibnamefont {Schug}},
  \bibinfo {author} {\bibfnamefont {P.}~\bibnamefont {Eich}}, \bibinfo {author}
  {\bibfnamefont {J.}~\bibnamefont {Huwer}}, \bibinfo {author} {\bibfnamefont
  {P.}~\bibnamefont {Müller}},\ and\ \bibinfo {author} {\bibfnamefont
  {J.}~\bibnamefont {Eschner}},\ }\bibfield  {title} {\bibinfo {title}
  {Experimental protocol for high-fidelity heralded photon-to-atom quantum
  state transfer},\ }\href {https://doi.org/10.1038/ncomms6527} {\bibfield
  {journal} {\bibinfo  {journal} {Nature Communications}\ }\textbf {\bibinfo
  {volume} {5}},\ \bibinfo {pages} {5527} (\bibinfo {year} {2014})}\BibitemShut
  {NoStop}%
\bibitem [{\citenamefont {Brekenfeld}\ \emph {et~al.}(2020)\citenamefont
  {Brekenfeld}, \citenamefont {Niemietz}, \citenamefont {Christesen},\ and\
  \citenamefont {Rempe}}]{brekenfeld2020quantum}%
  \BibitemOpen
  \bibfield  {author} {\bibinfo {author} {\bibfnamefont {M.}~\bibnamefont
  {Brekenfeld}}, \bibinfo {author} {\bibfnamefont {D.}~\bibnamefont
  {Niemietz}}, \bibinfo {author} {\bibfnamefont {J.~D.}\ \bibnamefont
  {Christesen}},\ and\ \bibinfo {author} {\bibfnamefont {G.}~\bibnamefont
  {Rempe}},\ }\bibfield  {title} {\bibinfo {title} {A quantum network node with
  crossed optical fibre cavities},\ }\href@noop {} {\bibfield  {journal}
  {\bibinfo  {journal} {Nature Physics}\ }\textbf {\bibinfo {volume} {16}},\
  \bibinfo {pages} {647} (\bibinfo {year} {2020})}\BibitemShut {NoStop}%
\bibitem [{\citenamefont {Kalb}\ \emph {et~al.}(2015)\citenamefont {Kalb},
  \citenamefont {Reiserer}, \citenamefont {Ritter},\ and\ \citenamefont
  {Rempe}}]{kalb_heralded_2015}%
  \BibitemOpen
  \bibfield  {author} {\bibinfo {author} {\bibfnamefont {N.}~\bibnamefont
  {Kalb}}, \bibinfo {author} {\bibfnamefont {A.}~\bibnamefont {Reiserer}},
  \bibinfo {author} {\bibfnamefont {S.}~\bibnamefont {Ritter}},\ and\ \bibinfo
  {author} {\bibfnamefont {G.}~\bibnamefont {Rempe}},\ }\bibfield  {title}
  {\bibinfo {title} {Heralded {Storage} of a {Photonic} {Quantum} {Bit} in a
  {Single} {Atom}},\ }\href {https://doi.org/10.1103/PhysRevLett.114.220501}
  {\bibfield  {journal} {\bibinfo  {journal} {Physical Review Letters}\
  }\textbf {\bibinfo {volume} {114}},\ \bibinfo {pages} {220501} (\bibinfo
  {year} {2015})}\BibitemShut {NoStop}%
\bibitem [{\citenamefont {Bhaskar}\ \emph {et~al.}(2020)\citenamefont
  {Bhaskar}, \citenamefont {Riedinger}, \citenamefont {Machielse},
  \citenamefont {Levonian}, \citenamefont {Nguyen}, \citenamefont {Knall},
  \citenamefont {Park}, \citenamefont {Englund}, \citenamefont {Lončar},
  \citenamefont {Sukachev},\ and\ \citenamefont
  {Lukin}}]{bhaskar_experimental_2020}%
  \BibitemOpen
  \bibfield  {author} {\bibinfo {author} {\bibfnamefont {M.~K.}\ \bibnamefont
  {Bhaskar}}, \bibinfo {author} {\bibfnamefont {R.}~\bibnamefont {Riedinger}},
  \bibinfo {author} {\bibfnamefont {B.}~\bibnamefont {Machielse}}, \bibinfo
  {author} {\bibfnamefont {D.~S.}\ \bibnamefont {Levonian}}, \bibinfo {author}
  {\bibfnamefont {C.~T.}\ \bibnamefont {Nguyen}}, \bibinfo {author}
  {\bibfnamefont {E.~N.}\ \bibnamefont {Knall}}, \bibinfo {author}
  {\bibfnamefont {H.}~\bibnamefont {Park}}, \bibinfo {author} {\bibfnamefont
  {D.}~\bibnamefont {Englund}}, \bibinfo {author} {\bibfnamefont
  {M.}~\bibnamefont {Lončar}}, \bibinfo {author} {\bibfnamefont {D.~D.}\
  \bibnamefont {Sukachev}},\ and\ \bibinfo {author} {\bibfnamefont {M.~D.}\
  \bibnamefont {Lukin}},\ }\bibfield  {title} {\bibinfo {title} {Experimental
  demonstration of memory-enhanced quantum communication},\ }\href
  {https://doi.org/10.1038/s41586-020-2103-5} {\bibfield  {journal} {\bibinfo
  {journal} {Nature}\ }\textbf {\bibinfo {volume} {580}},\ \bibinfo {pages}
  {60} (\bibinfo {year} {2020})}\BibitemShut {NoStop}%
\bibitem [{\citenamefont {Daiss}\ \emph {et~al.}(2021)\citenamefont {Daiss},
  \citenamefont {Langenfeld}, \citenamefont {Welte}, \citenamefont {Distante},
  \citenamefont {Thomas}, \citenamefont {Hartung}, \citenamefont {Morin},\ and\
  \citenamefont {Rempe}}]{daiss_quantum-logic_2021}%
  \BibitemOpen
  \bibfield  {author} {\bibinfo {author} {\bibfnamefont {S.}~\bibnamefont
  {Daiss}}, \bibinfo {author} {\bibfnamefont {S.}~\bibnamefont {Langenfeld}},
  \bibinfo {author} {\bibfnamefont {S.}~\bibnamefont {Welte}}, \bibinfo
  {author} {\bibfnamefont {E.}~\bibnamefont {Distante}}, \bibinfo {author}
  {\bibfnamefont {P.}~\bibnamefont {Thomas}}, \bibinfo {author} {\bibfnamefont
  {L.}~\bibnamefont {Hartung}}, \bibinfo {author} {\bibfnamefont
  {O.}~\bibnamefont {Morin}},\ and\ \bibinfo {author} {\bibfnamefont
  {G.}~\bibnamefont {Rempe}},\ }\bibfield  {title} {\bibinfo {title} {A
  quantum-logic gate between distant quantum-network modules},\ }\href
  {https://doi.org/10.1126/science.abe3150} {\bibfield  {journal} {\bibinfo
  {journal} {Science}\ }\textbf {\bibinfo {volume} {371}},\ \bibinfo {pages}
  {614} (\bibinfo {year} {2021})}\BibitemShut {NoStop}%
\bibitem [{\citenamefont {Knaut}\ \emph {et~al.}(2024)\citenamefont {Knaut},
  \citenamefont {Suleymanzade}, \citenamefont {Wei}, \citenamefont {Assumpcao},
  \citenamefont {Stas}, \citenamefont {Huan}, \citenamefont {Machielse},
  \citenamefont {Knall}, \citenamefont {Sutula}, \citenamefont {Baranes},
  \citenamefont {Sinclair}, \citenamefont {De-Eknamkul}, \citenamefont
  {Levonian}, \citenamefont {Bhaskar}, \citenamefont {Park}, \citenamefont
  {Lončar},\ and\ \citenamefont {Lukin}}]{knaut_entanglement_2024}%
  \BibitemOpen
  \bibfield  {author} {\bibinfo {author} {\bibfnamefont {C.~M.}\ \bibnamefont
  {Knaut}}, \bibinfo {author} {\bibfnamefont {A.}~\bibnamefont {Suleymanzade}},
  \bibinfo {author} {\bibfnamefont {Y.-C.}\ \bibnamefont {Wei}}, \bibinfo
  {author} {\bibfnamefont {D.~R.}\ \bibnamefont {Assumpcao}}, \bibinfo {author}
  {\bibfnamefont {P.-J.}\ \bibnamefont {Stas}}, \bibinfo {author}
  {\bibfnamefont {Y.~Q.}\ \bibnamefont {Huan}}, \bibinfo {author}
  {\bibfnamefont {B.}~\bibnamefont {Machielse}}, \bibinfo {author}
  {\bibfnamefont {E.~N.}\ \bibnamefont {Knall}}, \bibinfo {author}
  {\bibfnamefont {M.}~\bibnamefont {Sutula}}, \bibinfo {author} {\bibfnamefont
  {G.}~\bibnamefont {Baranes}}, \bibinfo {author} {\bibfnamefont
  {N.}~\bibnamefont {Sinclair}}, \bibinfo {author} {\bibfnamefont
  {C.}~\bibnamefont {De-Eknamkul}}, \bibinfo {author} {\bibfnamefont {D.~S.}\
  \bibnamefont {Levonian}}, \bibinfo {author} {\bibfnamefont {M.~K.}\
  \bibnamefont {Bhaskar}}, \bibinfo {author} {\bibfnamefont {H.}~\bibnamefont
  {Park}}, \bibinfo {author} {\bibfnamefont {M.}~\bibnamefont {Lončar}},\ and\
  \bibinfo {author} {\bibfnamefont {M.~D.}\ \bibnamefont {Lukin}},\ }\bibfield
  {title} {\bibinfo {title} {Entanglement of nanophotonic quantum memory nodes
  in a telecom network},\ }\href {https://doi.org/10.1038/s41586-024-07252-z}
  {\bibfield  {journal} {\bibinfo  {journal} {Nature}\ }\textbf {\bibinfo
  {volume} {629}},\ \bibinfo {pages} {573} (\bibinfo {year}
  {2024})}\BibitemShut {NoStop}%
\bibitem [{\citenamefont {Fleischhauer}\ \emph {et~al.}(2005)\citenamefont
  {Fleischhauer}, \citenamefont {Imamoglu},\ and\ \citenamefont
  {Marangos}}]{fleischhauer_electromagnetically_2005}%
  \BibitemOpen
  \bibfield  {author} {\bibinfo {author} {\bibfnamefont {M.}~\bibnamefont
  {Fleischhauer}}, \bibinfo {author} {\bibfnamefont {A.}~\bibnamefont
  {Imamoglu}},\ and\ \bibinfo {author} {\bibfnamefont {J.~P.}\ \bibnamefont
  {Marangos}},\ }\bibfield  {title} {\bibinfo {title} {Electromagnetically
  induced transparency: {Optics} in coherent media},\ }\href
  {https://doi.org/10.1103/RevModPhys.77.633} {\bibfield  {journal} {\bibinfo
  {journal} {Reviews of Modern Physics}\ }\textbf {\bibinfo {volume} {77}},\
  \bibinfo {pages} {633} (\bibinfo {year} {2005})}\BibitemShut {NoStop}%
\bibitem [{\citenamefont {Nunn}\ \emph {et~al.}(2007)\citenamefont {Nunn},
  \citenamefont {Walmsley}, \citenamefont {Raymer}, \citenamefont {Surmacz},
  \citenamefont {Waldermann}, \citenamefont {Wang},\ and\ \citenamefont
  {Jaksch}}]{nunn_mapping_2007}%
  \BibitemOpen
  \bibfield  {author} {\bibinfo {author} {\bibfnamefont {J.}~\bibnamefont
  {Nunn}}, \bibinfo {author} {\bibfnamefont {I.~A.}\ \bibnamefont {Walmsley}},
  \bibinfo {author} {\bibfnamefont {M.~G.}\ \bibnamefont {Raymer}}, \bibinfo
  {author} {\bibfnamefont {K.}~\bibnamefont {Surmacz}}, \bibinfo {author}
  {\bibfnamefont {F.~C.}\ \bibnamefont {Waldermann}}, \bibinfo {author}
  {\bibfnamefont {Z.}~\bibnamefont {Wang}},\ and\ \bibinfo {author}
  {\bibfnamefont {D.}~\bibnamefont {Jaksch}},\ }\bibfield  {title} {\bibinfo
  {title} {Mapping broadband single-photon wave packets into an atomic
  memory},\ }\href {https://doi.org/10.1103/PhysRevA.75.011401} {\bibfield
  {journal} {\bibinfo  {journal} {Physical Review A}\ }\textbf {\bibinfo
  {volume} {75}},\ \bibinfo {pages} {011401} (\bibinfo {year}
  {2007})}\BibitemShut {NoStop}%
\bibitem [{\citenamefont {Tanji}\ \emph {et~al.}(2009)\citenamefont {Tanji},
  \citenamefont {Ghosh}, \citenamefont {Simon}, \citenamefont {Bloom},\ and\
  \citenamefont {Vuletić}}]{tanji_heralded_2009}%
  \BibitemOpen
  \bibfield  {author} {\bibinfo {author} {\bibfnamefont {H.}~\bibnamefont
  {Tanji}}, \bibinfo {author} {\bibfnamefont {S.}~\bibnamefont {Ghosh}},
  \bibinfo {author} {\bibfnamefont {J.}~\bibnamefont {Simon}}, \bibinfo
  {author} {\bibfnamefont {B.}~\bibnamefont {Bloom}},\ and\ \bibinfo {author}
  {\bibfnamefont {V.}~\bibnamefont {Vuletić}},\ }\bibfield  {title} {\bibinfo
  {title} {Heralded {Single}-{Magnon} {Quantum} {Memory} for {Photon}
  {Polarization} {States}},\ }\href
  {https://doi.org/10.1103/PhysRevLett.103.043601} {\bibfield  {journal}
  {\bibinfo  {journal} {Physical Review Letters}\ }\textbf {\bibinfo {volume}
  {103}},\ \bibinfo {pages} {043601} (\bibinfo {year} {2009})}\BibitemShut
  {NoStop}%
\bibitem [{\citenamefont {Saffman}\ \emph {et~al.}(2010)\citenamefont
  {Saffman}, \citenamefont {Walker},\ and\ \citenamefont
  {Mølmer}}]{saffman_quantum_2010}%
  \BibitemOpen
  \bibfield  {author} {\bibinfo {author} {\bibfnamefont {M.}~\bibnamefont
  {Saffman}}, \bibinfo {author} {\bibfnamefont {T.~G.}\ \bibnamefont
  {Walker}},\ and\ \bibinfo {author} {\bibfnamefont {K.}~\bibnamefont
  {Mølmer}},\ }\bibfield  {title} {\bibinfo {title} {Quantum information with
  {Rydberg} atoms},\ }\href {https://doi.org/10.1103/RevModPhys.82.2313}
  {\bibfield  {journal} {\bibinfo  {journal} {Reviews of Modern Physics}\
  }\textbf {\bibinfo {volume} {82}},\ \bibinfo {pages} {2313} (\bibinfo {year}
  {2010})}\BibitemShut {NoStop}%
\bibitem [{\citenamefont {Kumlin}\ \emph {et~al.}(2023)\citenamefont {Kumlin},
  \citenamefont {Braun}, \citenamefont {Tresp}, \citenamefont {Stiesdal},
  \citenamefont {Hofferberth},\ and\ \citenamefont
  {Paris-Mandoki}}]{Kumlin_2023}%
  \BibitemOpen
  \bibfield  {author} {\bibinfo {author} {\bibfnamefont {J.}~\bibnamefont
  {Kumlin}}, \bibinfo {author} {\bibfnamefont {C.}~\bibnamefont {Braun}},
  \bibinfo {author} {\bibfnamefont {C.}~\bibnamefont {Tresp}}, \bibinfo
  {author} {\bibfnamefont {N.}~\bibnamefont {Stiesdal}}, \bibinfo {author}
  {\bibfnamefont {S.}~\bibnamefont {Hofferberth}},\ and\ \bibinfo {author}
  {\bibfnamefont {A.}~\bibnamefont {Paris-Mandoki}},\ }\bibfield  {title}
  {\bibinfo {title} {Quantum optics with {Rydberg} superatoms},\ }\href
  {https://doi.org/10.1088/2399-6528/acd51d} {\bibfield  {journal} {\bibinfo
  {journal} {Journal of Physics Communications}\ }\textbf {\bibinfo {volume}
  {7}},\ \bibinfo {pages} {052001} (\bibinfo {year} {2023})}\BibitemShut
  {NoStop}%
\bibitem [{\citenamefont {Pritchard}\ \emph {et~al.}(2010)\citenamefont
  {Pritchard}, \citenamefont {Maxwell}, \citenamefont {Gauguet}, \citenamefont
  {Weatherill}, \citenamefont {Jones},\ and\ \citenamefont
  {Adams}}]{PhysRevLett.105.193603}%
  \BibitemOpen
  \bibfield  {author} {\bibinfo {author} {\bibfnamefont {J.~D.}\ \bibnamefont
  {Pritchard}}, \bibinfo {author} {\bibfnamefont {D.}~\bibnamefont {Maxwell}},
  \bibinfo {author} {\bibfnamefont {A.}~\bibnamefont {Gauguet}}, \bibinfo
  {author} {\bibfnamefont {K.~J.}\ \bibnamefont {Weatherill}}, \bibinfo
  {author} {\bibfnamefont {M.~P.~A.}\ \bibnamefont {Jones}},\ and\ \bibinfo
  {author} {\bibfnamefont {C.~S.}\ \bibnamefont {Adams}},\ }\bibfield  {title}
  {\bibinfo {title} {Cooperative atom-light interaction in a blockaded
  {Rydberg} ensemble},\ }\href {https://doi.org/10.1103/PhysRevLett.105.193603}
  {\bibfield  {journal} {\bibinfo  {journal} {Physical Review Letters}\
  }\textbf {\bibinfo {volume} {105}},\ \bibinfo {pages} {193603} (\bibinfo
  {year} {2010})}\BibitemShut {NoStop}%
\bibitem [{\citenamefont {Petrosyan}\ \emph {et~al.}(2011)\citenamefont
  {Petrosyan}, \citenamefont {Otterbach},\ and\ \citenamefont
  {Fleischhauer}}]{PhysRevLett.107.213601}%
  \BibitemOpen
  \bibfield  {author} {\bibinfo {author} {\bibfnamefont {D.}~\bibnamefont
  {Petrosyan}}, \bibinfo {author} {\bibfnamefont {J.}~\bibnamefont
  {Otterbach}},\ and\ \bibinfo {author} {\bibfnamefont {M.}~\bibnamefont
  {Fleischhauer}},\ }\bibfield  {title} {\bibinfo {title} {Electromagnetically
  induced transparency with {Rydberg} atoms},\ }\href
  {https://doi.org/10.1103/PhysRevLett.107.213601} {\bibfield  {journal}
  {\bibinfo  {journal} {Physical Review Letters}\ }\textbf {\bibinfo {volume}
  {107}},\ \bibinfo {pages} {213601} (\bibinfo {year} {2011})}\BibitemShut
  {NoStop}%
\bibitem [{\citenamefont {Peyronel}\ \emph {et~al.}(2012)\citenamefont
  {Peyronel}, \citenamefont {Firstenberg}, \citenamefont {Liang}, \citenamefont
  {Hofferberth}, \citenamefont {Gorshkov}, \citenamefont {Pohl}, \citenamefont
  {Lukin},\ and\ \citenamefont {Vuleti{\'c}}}]{peyronel2012quantum}%
  \BibitemOpen
  \bibfield  {author} {\bibinfo {author} {\bibfnamefont {T.}~\bibnamefont
  {Peyronel}}, \bibinfo {author} {\bibfnamefont {O.}~\bibnamefont
  {Firstenberg}}, \bibinfo {author} {\bibfnamefont {Q.-Y.}\ \bibnamefont
  {Liang}}, \bibinfo {author} {\bibfnamefont {S.}~\bibnamefont {Hofferberth}},
  \bibinfo {author} {\bibfnamefont {A.~V.}\ \bibnamefont {Gorshkov}}, \bibinfo
  {author} {\bibfnamefont {T.}~\bibnamefont {Pohl}}, \bibinfo {author}
  {\bibfnamefont {M.~D.}\ \bibnamefont {Lukin}},\ and\ \bibinfo {author}
  {\bibfnamefont {V.}~\bibnamefont {Vuleti{\'c}}},\ }\bibfield  {title}
  {\bibinfo {title} {Quantum nonlinear optics with single photons enabled by
  strongly interacting atoms},\ }\href@noop {} {\bibfield  {journal} {\bibinfo
  {journal} {Nature}\ }\textbf {\bibinfo {volume} {488}},\ \bibinfo {pages}
  {57} (\bibinfo {year} {2012})}\BibitemShut {NoStop}%
\bibitem [{\citenamefont {Baur}\ \emph {et~al.}(2014)\citenamefont {Baur},
  \citenamefont {Tiarks}, \citenamefont {Rempe},\ and\ \citenamefont
  {D\"urr}}]{PhysRevLett.112.073901}%
  \BibitemOpen
  \bibfield  {author} {\bibinfo {author} {\bibfnamefont {S.}~\bibnamefont
  {Baur}}, \bibinfo {author} {\bibfnamefont {D.}~\bibnamefont {Tiarks}},
  \bibinfo {author} {\bibfnamefont {G.}~\bibnamefont {Rempe}},\ and\ \bibinfo
  {author} {\bibfnamefont {S.}~\bibnamefont {D\"urr}},\ }\bibfield  {title}
  {\bibinfo {title} {Single-photon switch based on {Rydberg} blockade},\ }\href
  {https://doi.org/10.1103/PhysRevLett.112.073901} {\bibfield  {journal}
  {\bibinfo  {journal} {Phys. Rev. Lett.}\ }\textbf {\bibinfo {volume} {112}},\
  \bibinfo {pages} {073901} (\bibinfo {year} {2014})}\BibitemShut {NoStop}%
\bibitem [{\citenamefont {Gorniaczyk}\ \emph {et~al.}(2014)\citenamefont
  {Gorniaczyk}, \citenamefont {Tresp}, \citenamefont {Schmidt}, \citenamefont
  {Fedder},\ and\ \citenamefont {Hofferberth}}]{PhysRevLett.113.053601}%
  \BibitemOpen
  \bibfield  {author} {\bibinfo {author} {\bibfnamefont {H.}~\bibnamefont
  {Gorniaczyk}}, \bibinfo {author} {\bibfnamefont {C.}~\bibnamefont {Tresp}},
  \bibinfo {author} {\bibfnamefont {J.}~\bibnamefont {Schmidt}}, \bibinfo
  {author} {\bibfnamefont {H.}~\bibnamefont {Fedder}},\ and\ \bibinfo {author}
  {\bibfnamefont {S.}~\bibnamefont {Hofferberth}},\ }\bibfield  {title}
  {\bibinfo {title} {Single-photon transistor mediated by interstate {Rydberg}
  interactions},\ }\href {https://doi.org/10.1103/PhysRevLett.113.053601}
  {\bibfield  {journal} {\bibinfo  {journal} {Phys. Rev. Lett.}\ }\textbf
  {\bibinfo {volume} {113}},\ \bibinfo {pages} {053601} (\bibinfo {year}
  {2014})}\BibitemShut {NoStop}%
\bibitem [{\citenamefont {Tiarks}\ \emph {et~al.}(2014)\citenamefont {Tiarks},
  \citenamefont {Baur}, \citenamefont {Schneider}, \citenamefont {D\"urr},\
  and\ \citenamefont {Rempe}}]{PhysRevLett.113.053602}%
  \BibitemOpen
  \bibfield  {author} {\bibinfo {author} {\bibfnamefont {D.}~\bibnamefont
  {Tiarks}}, \bibinfo {author} {\bibfnamefont {S.}~\bibnamefont {Baur}},
  \bibinfo {author} {\bibfnamefont {K.}~\bibnamefont {Schneider}}, \bibinfo
  {author} {\bibfnamefont {S.}~\bibnamefont {D\"urr}},\ and\ \bibinfo {author}
  {\bibfnamefont {G.}~\bibnamefont {Rempe}},\ }\bibfield  {title} {\bibinfo
  {title} {Single-photon transistor using a f\"orster resonance},\ }\href
  {https://doi.org/10.1103/PhysRevLett.113.053602} {\bibfield  {journal}
  {\bibinfo  {journal} {Phys. Rev. Lett.}\ }\textbf {\bibinfo {volume} {113}},\
  \bibinfo {pages} {053602} (\bibinfo {year} {2014})}\BibitemShut {NoStop}%
\bibitem [{\citenamefont {Tiarks}\ \emph {et~al.}(2019)\citenamefont {Tiarks},
  \citenamefont {Schmidt-Eberle}, \citenamefont {Stolz}, \citenamefont
  {Rempe},\ and\ \citenamefont {D{\"u}rr}}]{tiarks2019photon}%
  \BibitemOpen
  \bibfield  {author} {\bibinfo {author} {\bibfnamefont {D.}~\bibnamefont
  {Tiarks}}, \bibinfo {author} {\bibfnamefont {S.}~\bibnamefont
  {Schmidt-Eberle}}, \bibinfo {author} {\bibfnamefont {T.}~\bibnamefont
  {Stolz}}, \bibinfo {author} {\bibfnamefont {G.}~\bibnamefont {Rempe}},\ and\
  \bibinfo {author} {\bibfnamefont {S.}~\bibnamefont {D{\"u}rr}},\ }\bibfield
  {title} {\bibinfo {title} {A photon--photon quantum gate based on {Rydberg}
  interactions},\ }\href@noop {} {\bibfield  {journal} {\bibinfo  {journal}
  {Nature Physics}\ }\textbf {\bibinfo {volume} {15}},\ \bibinfo {pages} {124}
  (\bibinfo {year} {2019})}\BibitemShut {NoStop}%
\bibitem [{\citenamefont {Vaneecloo}\ \emph {et~al.}(2022)\citenamefont
  {Vaneecloo}, \citenamefont {Garcia},\ and\ \citenamefont
  {Ourjoumtsev}}]{vaneecloo2022intracavity}%
  \BibitemOpen
  \bibfield  {author} {\bibinfo {author} {\bibfnamefont {J.}~\bibnamefont
  {Vaneecloo}}, \bibinfo {author} {\bibfnamefont {S.}~\bibnamefont {Garcia}},\
  and\ \bibinfo {author} {\bibfnamefont {A.}~\bibnamefont {Ourjoumtsev}},\
  }\bibfield  {title} {\bibinfo {title} {Intracavity {Rydberg} superatom for
  optical quantum engineering: coherent control, single-shot detection, and
  optical $\pi$ phase shift},\ }\href@noop {} {\bibfield  {journal} {\bibinfo
  {journal} {Physical Review X}\ }\textbf {\bibinfo {volume} {12}},\ \bibinfo
  {pages} {021034} (\bibinfo {year} {2022})}\BibitemShut {NoStop}%
\bibitem [{\citenamefont {Stolz}\ \emph {et~al.}(2022)\citenamefont {Stolz},
  \citenamefont {Hegels}, \citenamefont {Winter}, \citenamefont {R{\"o}hr},
  \citenamefont {Hsiao}, \citenamefont {Husel}, \citenamefont {Rempe},\ and\
  \citenamefont {D{\"u}rr}}]{stolz2022quantum}%
  \BibitemOpen
  \bibfield  {author} {\bibinfo {author} {\bibfnamefont {T.}~\bibnamefont
  {Stolz}}, \bibinfo {author} {\bibfnamefont {H.}~\bibnamefont {Hegels}},
  \bibinfo {author} {\bibfnamefont {M.}~\bibnamefont {Winter}}, \bibinfo
  {author} {\bibfnamefont {B.}~\bibnamefont {R{\"o}hr}}, \bibinfo {author}
  {\bibfnamefont {Y.-F.}\ \bibnamefont {Hsiao}}, \bibinfo {author}
  {\bibfnamefont {L.}~\bibnamefont {Husel}}, \bibinfo {author} {\bibfnamefont
  {G.}~\bibnamefont {Rempe}},\ and\ \bibinfo {author} {\bibfnamefont
  {S.}~\bibnamefont {D{\"u}rr}},\ }\bibfield  {title} {\bibinfo {title}
  {Quantum-logic gate between two optical photons with an average efficiency
  above 40\%},\ }\href@noop {} {\bibfield  {journal} {\bibinfo  {journal}
  {Physical Review X}\ }\textbf {\bibinfo {volume} {12}},\ \bibinfo {pages}
  {021035} (\bibinfo {year} {2022})}\BibitemShut {NoStop}%
\bibitem [{\citenamefont {Sun}\ \emph {et~al.}(2022)\citenamefont {Sun},
  \citenamefont {Yu}, \citenamefont {An}, \citenamefont {Li}, \citenamefont
  {Yang}, \citenamefont {Bao},\ and\ \citenamefont
  {Pan}}]{sun_deterministic_2022}%
  \BibitemOpen
  \bibfield  {author} {\bibinfo {author} {\bibfnamefont {P.-F.}\ \bibnamefont
  {Sun}}, \bibinfo {author} {\bibfnamefont {Y.}~\bibnamefont {Yu}}, \bibinfo
  {author} {\bibfnamefont {Z.-Y.}\ \bibnamefont {An}}, \bibinfo {author}
  {\bibfnamefont {J.}~\bibnamefont {Li}}, \bibinfo {author} {\bibfnamefont
  {C.-W.}\ \bibnamefont {Yang}}, \bibinfo {author} {\bibfnamefont {X.-H.}\
  \bibnamefont {Bao}},\ and\ \bibinfo {author} {\bibfnamefont {J.-W.}\
  \bibnamefont {Pan}},\ }\bibfield  {title} {\bibinfo {title} {Deterministic
  {Time}-{Bin} {Entanglement} between a {Single} {Photon} and an {Atomic}
  {Ensemble}},\ }\href {https://doi.org/10.1103/PhysRevLett.128.060502}
  {\bibfield  {journal} {\bibinfo  {journal} {Physical Review Letters}\
  }\textbf {\bibinfo {volume} {128}},\ \bibinfo {pages} {060502} (\bibinfo
  {year} {2022})}\BibitemShut {NoStop}%
\bibitem [{\citenamefont {Yang}\ \emph {et~al.}(2022)\citenamefont {Yang},
  \citenamefont {Yu}, \citenamefont {Li}, \citenamefont {Jing}, \citenamefont
  {Bao},\ and\ \citenamefont {Pan}}]{yang_sequential_2022}%
  \BibitemOpen
  \bibfield  {author} {\bibinfo {author} {\bibfnamefont {C.-W.}\ \bibnamefont
  {Yang}}, \bibinfo {author} {\bibfnamefont {Y.}~\bibnamefont {Yu}}, \bibinfo
  {author} {\bibfnamefont {J.}~\bibnamefont {Li}}, \bibinfo {author}
  {\bibfnamefont {B.}~\bibnamefont {Jing}}, \bibinfo {author} {\bibfnamefont
  {X.-H.}\ \bibnamefont {Bao}},\ and\ \bibinfo {author} {\bibfnamefont {J.-W.}\
  \bibnamefont {Pan}},\ }\bibfield  {title} {\bibinfo {title} {Sequential
  generation of multiphoton entanglement with a {Rydberg} superatom},\ }\href
  {https://doi.org/10.1038/s41566-022-01054-3} {\bibfield  {journal} {\bibinfo
  {journal} {Nature Photonics}\ }\textbf {\bibinfo {volume} {16}},\ \bibinfo
  {pages} {658} (\bibinfo {year} {2022})}\BibitemShut {NoStop}%
\bibitem [{\citenamefont {O'Brien}(2007)}]{obrien_optical_2007}%
  \BibitemOpen
  \bibfield  {author} {\bibinfo {author} {\bibfnamefont {J.~L.}\ \bibnamefont
  {O'Brien}},\ }\bibfield  {title} {\bibinfo {title} {Optical {Quantum}
  {Computing}},\ }\href {https://doi.org/10.1126/science.1142892} {\bibfield
  {journal} {\bibinfo  {journal} {Science}\ }\textbf {\bibinfo {volume}
  {318}},\ \bibinfo {pages} {1567} (\bibinfo {year} {2007})}\BibitemShut
  {NoStop}%
\bibitem [{\citenamefont {Xu}\ \emph {et~al.}(2020)\citenamefont {Xu},
  \citenamefont {Ma}, \citenamefont {Zhang}, \citenamefont {Lo},\ and\
  \citenamefont {Pan}}]{xu_secure_2020}%
  \BibitemOpen
  \bibfield  {author} {\bibinfo {author} {\bibfnamefont {F.}~\bibnamefont
  {Xu}}, \bibinfo {author} {\bibfnamefont {X.}~\bibnamefont {Ma}}, \bibinfo
  {author} {\bibfnamefont {Q.}~\bibnamefont {Zhang}}, \bibinfo {author}
  {\bibfnamefont {H.-K.}\ \bibnamefont {Lo}},\ and\ \bibinfo {author}
  {\bibfnamefont {J.-W.}\ \bibnamefont {Pan}},\ }\bibfield  {title} {\bibinfo
  {title} {Secure quantum key distribution with realistic devices},\ }\href
  {https://doi.org/10.1103/RevModPhys.92.025002} {\bibfield  {journal}
  {\bibinfo  {journal} {Reviews of Modern Physics}\ }\textbf {\bibinfo {volume}
  {92}},\ \bibinfo {pages} {025002} (\bibinfo {year} {2020})}\BibitemShut
  {NoStop}%
\bibitem [{\citenamefont {Dudin}\ and\ \citenamefont
  {Kuzmich}(2012)}]{2012Strongly}%
  \BibitemOpen
  \bibfield  {author} {\bibinfo {author} {\bibfnamefont {Y.~O.}\ \bibnamefont
  {Dudin}}\ and\ \bibinfo {author} {\bibfnamefont {A.}~\bibnamefont
  {Kuzmich}},\ }\bibfield  {title} {\bibinfo {title} {Strongly interacting
  {Rydberg} excitations of a cold atomic gas},\ }\href@noop {} {\bibfield
  {journal} {\bibinfo  {journal} {Science}\ }\textbf {\bibinfo {volume}
  {336}},\ \bibinfo {pages} {887} (\bibinfo {year} {2012})}\BibitemShut
  {NoStop}%
\bibitem [{\citenamefont {Greenbaum}(2015)}]{greenbaum2015introduction}%
  \BibitemOpen
  \bibfield  {author} {\bibinfo {author} {\bibfnamefont {D.}~\bibnamefont
  {Greenbaum}},\ }\bibfield  {title} {\bibinfo {title} {Introduction to quantum
  gate set tomography},\ }\href@noop {} {\bibfield  {journal} {\bibinfo
  {journal} {arXiv preprint arXiv:1509.02921}\ } (\bibinfo {year}
  {2015})}\BibitemShut {NoStop}%
\bibitem [{\citenamefont {Beterov}\ \emph {et~al.}(2016)\citenamefont
  {Beterov}, \citenamefont {Saffman}, \citenamefont {Yakshina}, \citenamefont
  {Tretyakov}, \citenamefont {Entin}, \citenamefont {Hamzina},\ and\
  \citenamefont {Ryabtsev}}]{beterov2016simulated}%
  \BibitemOpen
  \bibfield  {author} {\bibinfo {author} {\bibfnamefont {I.}~\bibnamefont
  {Beterov}}, \bibinfo {author} {\bibfnamefont {M.}~\bibnamefont {Saffman}},
  \bibinfo {author} {\bibfnamefont {E.}~\bibnamefont {Yakshina}}, \bibinfo
  {author} {\bibfnamefont {D.}~\bibnamefont {Tretyakov}}, \bibinfo {author}
  {\bibfnamefont {V.}~\bibnamefont {Entin}}, \bibinfo {author} {\bibfnamefont
  {G.}~\bibnamefont {Hamzina}},\ and\ \bibinfo {author} {\bibfnamefont
  {I.}~\bibnamefont {Ryabtsev}},\ }\bibfield  {title} {\bibinfo {title}
  {Simulated quantum process tomography of quantum gates with {Rydberg}
  superatoms},\ }\href@noop {} {\bibfield  {journal} {\bibinfo  {journal}
  {Journal of Physics B: Atomic, Molecular and Optical Physics}\ }\textbf
  {\bibinfo {volume} {49}},\ \bibinfo {pages} {114007} (\bibinfo {year}
  {2016})}\BibitemShut {NoStop}%
\end{thebibliography}%

\end{document}